# Momentum-dependent oscillator strength crossover of excitons and plasmons in two-dimensional PtSe$_2$

Jinhua Hong[1,#], Mark Kamper Svendsen[2,#,*], Masanori Koshino[1], Thomas Pichler[3], Hua Xu[4], Kazu Suenaga[1,6,*], Kristian Sommer Thygesen[2,5]

[1]Nanomaterials Research Institute, National Institute of Advanced Industrial Science and Technology (AIST), Tsukuba 305-8565, Japan

[2]Computational Atomic-scale Materials Design (CAMD), Department of Physics, Technical University of Denmark, Kongens Lyngby, Denmark

[3]Faculty of Physics, University of Vienna, Strudlhofgasse 4, A-1090 Vienna, Austria

[4]Key Laboratory of Applied Surface and Colloid Chemistry, School of Materials Science and Engineering, Shaanxi Normal University, Xi'an 710119, P. R. China

[5]Center for Nanostructured Graphene (CNG), Department of Physics, Technical University of Denmark, Kongens Lyngby, Denmark

[6]The Institute of Scientific and Industrial Research (ISIR-SANKEN), Osaka University, Ibaraki 567-0047, Japan

[#]These authors contributed to this work equally.

[*]E-mail: suenaga-kazu@sanken.osaka-u.ac.jp, markas@dtu.dk

**The 1T-phase layered PtX$_2$ chalcogenides has attracted widespread interest due to its thickness dependent metal-semiconductor transition driven by strong interlayer coupling. While the ground state properties of this paradigmatic material system have been widely explored, its fundamental excitation spectrum remains poorly understood. Here we combine first principles calculations with momentum ($q$) resolved electron energy loss spectroscopy ($q$-EELS) to study the collective excitations in 1T-PtSe$_2$ from the monolayer limit to the bulk. At finite momentum transfer all the spectra are dominated by two distinct interband plasmons that disperse to higher energy with increasing $q$. Interestingly, the absence of long-range screening in the two-dimensional (2D) limit, inhibits the**

**formation of long wavelength plasmons. Consequently, in the small-*q* limit, excitations in monolayer PtSe$_2$ are exclusively of excitonic nature, and the loss spectrum coincides with the optical spectrum. Our work unravels the excited state spectrum of layered 1T-PtSe$_2$ and establishes the qualitatively different momentum dependence of excitons and plasmons in 2D materials.**

## Introduction

Platinum dichalcogenides (PtX$_2$, X=S, Se, Te) are emerging as a new class of two-dimensional (2D) layered materials[1, 2] with thickness- and anion-dependent electronic properties resulting from the strong interlayer coupling[3, 4, 5]. In this octahedral-coordinated 1T-phase system, PtSe$_2$ is an indirect semiconductor in monolayer and bilayer forms and evolves into a metal as the thickness increases to trilayer and towards the bulk[4, 6]. This semiconductor-to-metal transition is caused by the giant coupling of *p$_z$-p$_z$* states of two adjacent interlayer chalcogen atoms[4, 5]. This is in contrast to the well-known 2H-MoS$_2$ system where the weaker interlayer coupling reduces the gap as more layers are stacked, but is insufficient to drive a semiconductor-to-metal transition. When the chalcogen changes from S over Se to Te, the bulk PtX$_2$ turns from a semiconductor (PtS$_2$) to a metal (PtSe$_2$ and PtTe$_2$). Chia et al[7] showed that the enhanced conductivity across the chalcogen group gives rise to a monotonic increasing catalytic performance in the hydrogen evolution reaction[8]. Vacancy defect induced magnetism and a layer-modulated ferromagnetic-to-antiferromagnetic crossover have been observed in both metallic PtSe$_2$ few-layers[9] and semiconducting bi- and mono-layers[10]. In the electronic transport characteristics, an *n*- to *p*-type conductivity modulation has been realized in vacancy doped monolayer PtSe$_2$ via controlled chemical vapor deposition[11]. The monolayer PtX$_2$ semiconductors present high electron mobilities and are stable in air[4], demonstrating their potential for mid-infrared optoelectronics[12], magnetism, and catalysis applications.

While the transport and catalytic properties of PtX$_2$ have been thoroughly investigated, its electronic excitation spectrum, including the collective excitations (excitons and plasmons), remains largely unexplored. Understanding these excitations[13, 14, 15] is the key to controlling the optical properties of the material and could ultimately pave the way to photonic or optoelectronic applications[12, 16, 17, 18, 19]. The optical excitations in the $MoX_2$ and $WX_2$ semiconducting TMDs have been intensely studied during the past years because of the unique exciton physics found in their mono- and few-layer structures[20, 21]. In comparison, the optical properties of few-layer $PtX_2$ have only recently gained attention, and only the most basic properties of their excitons have so far been addressed[22, 23, 24, 25]. The plasmon excitations of bulk and thin layers of $PtTe_2$ were reported by Ghosh *et al*[26]. However, in these measurements the samples were not atomically thin and substrate-sample interactions could not be ruled out. To date, the electronic excitation spectrum of freestanding few-layer $PtX_2$ remains unexplored, and it is unclear how it depends on the number of layers and the metal-semiconductor transition taking place as the thickness decreases to the monolayer limit.

In this work, we employ momentum ($q$) resolved electron energy loss spectroscopy ($q$-EELS)[27, 28, 29, 30] in a transmission electron microscope (TEM) to probe the valence excitations of $PtSe_2$ samples of different thicknesses. In the high energy range, we identify the existence of in-plane polarized plasmons at 7~10 eV across various film thicknesses. By tracking their $q$-dispersion we find that this plasmon excitation is enhanced at large nonzero $q$, similar to the π plasmon of graphene. At lower energies, a peak at ~2eV appears and dominates the EEL spectrum, particularly for the thinner samples. We ascribe this peak to excitonic effects in the two-dimensional samples and show that it is highly sensitive to $q$. Meanwhile, a new plasmon-like feature emerges between the exciton and the high-energy plasmon, which has not been observed previously in other TMDs. The loss functions calculated from first principles within the random phase approximation (RPA) confirm the plasmonic nature of the high- and intermediate-energy peaks. We furthermore employ the Bethe Salpeter Equation (BSE) to prove the excitonic nature of the low energy peak in the monolayer spectrum and map out the *k*-space wave function of the lowest exciton. Overall, we find good agreement between theory and experiment, which allows us to perform a detailed study of the collective excitations in layered PtSe$_2$ and elucidate the intricate connection between sample dimensionality (2D *vs* 3D), dielectric screening, and the relative importance of excitons and plasmons as a function of momentum transfer, *q*.

## Results

### Sample fabrication and *q*-EELS setup

High quality monolayer and few-layer PtSe$_2$ are grown on mica substrates by direct selenization of PtCl$_4$ in a chemical vapor deposition (CVD) furnace at 700 °C. Single-crystal domains are easily formed with a typical size of tens of micrometers. The as-grown samples are transferred onto TEM grids using PMMA spin coating and KI etching, and then washed in acetone and isopropanol to remove the surface residues. Before the *q*-EELS experiments in the TEM, the PtSe$_2$ samples are annealed at 200 °C for 3 hours. All the *q*-EEL spectra are collected in a diffraction mode using parallel beam illumination, with an energy resolution of 40 meV and a momentum resolution of 0.03 Å$^{-1}$.

### Atomic and electronic structures of PtSe$_2$

Figure 1 shows the atomic and electronic band structures of layered 1T-PtSe$_2$. In the structural models in Fig.1a, 1b, the Pt atoms (in cyan) are coordinated with 6 Se atoms (in yellow), forming an octahedral crystal field with $d^2sp^3$ hybridization. This is the known 1T phase structure, and each layer is stacked AA′ in order to form the bulk system. In Fig. 1c, the atomically resolved annular dark-field scanning transmission electron microscopy (ADF-STEM) image clearly shows the atomic coordination of monolayer PtSe$_2$ in the 1T phase. The brighter Pt atom is surrounded by 6 darker Se atoms. Based on this atomic structure, our density functional theory (DFT) calculations provide the single-particle electronic band structure of PtSe$_2$ as a function of film thickness (Fig. 1d). In the monolayer case, PtSe$_2$ is an indirect semiconductor

($E_g$ = 1.1 eV) with the valence band maximum (VBM) at the Γ point and the conduction band minimum (CBM) between Γ and M. As the thickness increases to bilayer, the bandgap decreases drastically to only 0.2 eV. When further increasing to trilayer or bulk, the electronic structure evolves into a metal. Figure 1d shows the metal-to-semiconductor transition as the thickness decreases from bulk to monolayer (also Fig. S1). This transition originates from the strong electronic interlayer hybridization, which occurs because the wave functions of the valence and conduction bands have a strong component of Se-$p_z$ orbitals (see Fig. S1 in the supplementary material). This is in contrast to the well-known indirect-to-direct gap transition in 2H-MoS$_2$ where the hybridization is significantly weaker due to the wave functions of the valence and conduction bands being mainly localized on the inner Mo atoms.

**Loss functions of PtSe$_2$ for different thicknesses**

To explore the electronic excitations of PtSe$_2$, we employ both $q$-EELS measurements and first principles calculations. Figure 2a-c shows the low loss fine structure of single-layer (1L), 4-layer (4L), and bulk PtSe$_2$ acquired at different $q$ along the ΓM direction (assuming that in-plane anisotropy can be ignored within the limited $q$ range < 0.2 Å$^{-1}$). Three clear peaks denoted A, B, and C dominate the electronic excitation spectrum. Interestingly, the relative intensities of the three peaks vary strongly with both $q$ and layer thickness. Understanding the origin of these variations and the nature of the electronic excitations behind the peaks (whether excitonic or plasmonic), is a main focus of this work.

The high-energy C peak at ~ 8 eV and the intermediate B peak at ~ 4 eV both have positive and non-linear dispersion with $q$, see Figure 3a. For all thicknesses these peaks dominate the EEL spectra for large $q$. For smaller $q$, the C peak disappears almost entirely for all thicknesses while the B peak remains in the bulk spectrum, becomes weaker in 4L and disappears in the monolayer limit. The fact that the B and C spectral features are well reproduced by our random phase approximation (RPA) calculations, see Fig. 2d-f, suggests that the underlying excitations are of plasmonic nature. This is because the RPA neglects the attractive electron-hole (e-h) interaction responsible for exciton formation and only include e-h exchange that drives plasmon formation.

For momentum-transfer below 0.05 Å$^{-1}$ a low-lying peak A at ~ 2 eV appears in the EEL spectrum of all samples. Interestingly, the A peak is only present at small $q$ and becomes dwarfed by the B and C plasmonic features for larger $q$ - a trend that is particularly pronounced in the monolayer limit. In the following we discuss why this observation is indicative of the peak A being of excitonic nature in the two dimensional samples.[23]

**$q$ dependence of the electronic screening in 2D *vs* 3D**

To understand why the small-$q$ EEL spectrum of the monolayer is dominated by the excitonic A peak while the high-energy B and C plasmon peaks dominate for larger $q$, we consider the Bethe Salpeter equation (BSE) which can account for both single-particle excitations, plasmons and excitons[31, 32, 33]. In the BSE, the excitations are found as solutions to the eigenvalue problem,

$$\sum_{SS'} H_{SS'}(\boldsymbol{q}) F_{S'}^{\lambda}(\boldsymbol{q}) = E^{\lambda} F_S(\boldsymbol{q}) \qquad (1)$$

where $H_{SS'}(\boldsymbol{q})$ is the two-particle BSE Hamiltonian in the e-h pair orbital basis labeled by the pair-orbital index $S = (n, m, \boldsymbol{k}, \boldsymbol{q})$, $F_{S'}^{\lambda}(\boldsymbol{q})$ is the exciton wave function of the $\lambda^{\text{th}}$ solution and $E^{\lambda}$ is the corresponding energy. The BSE Hamiltonian reads,

$$H_{SS'}(\boldsymbol{q}) = \left(\epsilon_{m,\boldsymbol{k}+\boldsymbol{q}}^{QP} - \epsilon_{n,\boldsymbol{k}}^{QP}\right)\delta_{SS'} - \left(f_{m,\boldsymbol{k}+\boldsymbol{q}} - f_{n,\boldsymbol{k}}\right) K_{SS'}(\boldsymbol{q}) \qquad (2)$$

The diagonal holds the quasiparticle (QP) energies of the electron and hole state. The exchange-correlation-kernel, $K_{SS'}(\boldsymbol{q})$, that renormalizes and mixes the single-particle transitions, can be divided into two constituent parts: the e-h exchange interaction ($V$) and the direct screened interaction ($W$),

$$K_{SS'}(\boldsymbol{q}) = 2V_{SS'}(\boldsymbol{q}) - W_{SS'}(\boldsymbol{q}) \qquad (3)$$

where the factor 2 appears because we have specialized to singlet excitations. Setting $K=0$, yields the single-particle spectrum of bare e-h transitions while setting $W=0$ yields the RPA. In general, both $V$ and $W$ are positive, and thus the former increases the energy of the collective excitation while the latter decreases its energy, relative to the bare QP transition energies of the constituent e-h pairs. This leads to the observation (which may also be taken as a definition) that $V$ is responsible for plasmon formation and $W$ is responsible for exciton formation. The relative importance of these two terms therefore dictates what type of excitations will dominate at a given $\boldsymbol{q}$. We note that in this discussion we do not distinguish between interband and intraband plasmons.

The exchange interaction takes the form,

$$V_{SS'}(\boldsymbol{q}) = \int d\boldsymbol{r}d\boldsymbol{r}' \frac{\psi_{nk}(\boldsymbol{r})\psi^*_{mk+q}(\boldsymbol{r})\psi^*_{n'k'}(\boldsymbol{r}')\psi_{m'k'+q}(\boldsymbol{r}')}{|\boldsymbol{r}-\boldsymbol{r}'|} \qquad (4)$$

It can be shown that for small momentum transfers, the long wavelength (small $q$) Fourier component of $\psi_{nk}(\boldsymbol{r})\psi^*_{mk+q}(\boldsymbol{r})$ is proportional to $q$ regardless of the dimensionality of the system. In two dimensions, the Fourier transformed Coulomb interaction goes as $v_q \sim \frac{1}{q}$ for small $q$, which means that the exchange interaction goes as $V_{SS'}(q) \sim q$ and therefore it becomes negligible in the optical limit ($q \to 0$). For this reason, we expect very weak plasmon formation at small $q$ in atomically thin 2D materials. This is in stark contrast to the 3D case where the Coulomb interaction goes as $v_q \sim \frac{1}{q^2}$ meaning that the exchange interaction in 3D can be significant also at small $q$. In the 3D case, plasmons can therefore exist even in the optical limit, and this represents a fundamental qualitative difference between plasmons in 2D and 3D.

Unlike the exchange interaction matrix elements, the direct screened interaction matrix elements $W_{SS'}(\boldsymbol{q})$ remain finite and more or less independent of $q$ in the optical limit. This means that, in contrast to plasmons, excitons can still form in 2D materials for small $q$. In fact, in the limit of small $q$, excitonic effects are extremely pronounced in 2D semiconductors due to the weak screening and can greatly reconstruct the non-interacting spectrum (see Fig. 5 and related discussion). However, as $q$ increases, the importance of the exchange interaction grows to dominate the direct interaction, leading to an excitation spectrum dominated primarily by plasmons. As demonstrated in Fig. 2, the excitonic peak A dominates at small $q$ while the B and C plasmonic peaks grow dominant as $q$ increases. To the best of our knowledge, this is the first direct observation of the intricate momentum-dependent difference between exciton and plasmon formation in a 2D material.

By tracking the peak energies (Fig. 2a-c) as a function of momentum $q$, we obtain the dispersion of the plasmon peaks B and C, as shown in Fig. 3a. Both the B and C peaks exhibit positive dispersion for all samples thicknesses. For $q > 0.04$ Å$^{-1}$, the dispersion of both peaks is $\sqrt{q}$-like for all material thicknesses. For the monolayer, this is well reproduced by our RPA calculations (Fig 3b) while the agreement with theory is less impressive for the 4L and bulk systems (Fig. S2) where RPA does predict the plasmon energies correctly to within 0.5 eV, but yields a too flat dispersion. In the small-$q$ range, the experimental peak energies flatten out and deviate from the $\sqrt{q}$-dispersion. We note that, due to the limited momentum resolution in the diffraction mode, the spectra at $q < 0.04$ Å$^{-1}$ are influenced by the excitation at the optical limit $q\rightarrow 0$, which may affect the dispersion curve in this $q$ range. We further observe that the B and C peaks blue shift with sample thickness, which is also reproduced by our RPA calculations, see Fig. S2.

**Eigenmode analysis of the plasmon excitations**

In general, the $q$-EEL spectrum is directly related to the inverse of the macroscopic dielectric function, $\epsilon_M(q,\omega)$. While $\epsilon_M$ is thus the relevant physical observable, it provides little insight into the nature of the excitations underlying the different spectral features, e.g. whether they are of single-particle, plasmonic or excitonic character, whether one or several different excitations contribute to a specific spectral feature, or what the spatial form and symmetry of such excitations are. As already discussed in relation to Eq. (3), the first question is largely a matter of which level of theory is required to reproduce a given spectral feature. At this point we have established that the B and C peaks are well described at the RPA level and thus have single-particle and/or plasmonic nature. To investigate the plasmonic nature of the B and C peaks in more detail and to clarify whether they represent one, two, or several different plasmon excitations, we turn to an eigenmode decomposition of the dielectric function. Following Andersen *et al*[34], the microscopic dielectric function can be decomposed into a set of eigenmodes,

$$\epsilon(\mathbf{r},\mathbf{r}',\omega) = \sum_n \epsilon_n(\omega)\phi_n(\mathbf{r},\omega)\rho_n(\mathbf{r}',\omega) \qquad (5)$$

where $\epsilon_n(\omega)$ is the eigenvalue of the $n^{\text{th}}$ mode of the dielectric function and $\phi_n(\mathbf{r},\omega)$ and $\rho_n(\mathbf{r}',\omega)$ are the left and right eigenvectors which satisfy the Poisson equation $\nabla^2 \phi_n = -4\pi\rho_n$. Here $\phi_n$ is the potential and $\rho_n$ is the induced density of the $n^{\text{th}}$ eigenmode. The modes satisfying $\text{Re}\epsilon_n(\omega) = 0$ for some frequency $\omega = \omega_p$ are plasmonic as they represent self-sustained charge density oscillations. We note that only modes with a spatially symmetric profile (described by $\rho_n(\mathbf{r},\omega)$) perpendicular to the PtSe$_2$ film, will contribute to the macroscopic in-plane $q$ EELS.

Our spectral analysis shows that the full RPA EEL spectra for all material thicknesses and in the relevant $(q,\omega)$-range, can be almost entirely accounted for by two dielectric eigenmodes. Figure 4 shows the individual contribution of the two eigenvalues to $\text{Im}\epsilon(\omega)$, $\text{Re}\epsilon(\omega)$, and $\text{Im}\epsilon^{-1}(\omega)$ for the monolayer (left) and bulk (right) for momentum transfers $q$=0.04 Å$^{-1}$ and $q$=0.12 Å$^{-1}$, respectively. For the monolayer, we see that the two modes are separately responsible for the B and C peaks, i.e. each peak can be associated with a distinct excitation. As $q$ increases both peaks evolve from single-particle interband transition peaks into true plasmons, as defined by a zero-crossing of $\text{Re}\epsilon_n(\omega)$. For the bulk system, we see that one mode is entirely responsible for the B peak. This mode also provides the dominant contribution to the C peak but there is a contribution from one additional mode. The dielectric modes of the bulk system vary much less with $q$ and are plasmonic for all values of $q$. We ascribe the stronger plasmonic character of the excitations in bulk versus 1L, to the stronger dielectric screening response in 3D compared to 2D. In particular, as previously discussed, for small values of $q$ the e-h exchange term, i.e. the $V$ in Eq. (3) that is responsible for screening and drives plasmon formation, is directly proportional to $q$. This point becomes particularly clear in the optical limit ($q$=0) where the long range dielectric screening vanishes identically in 2D. Fig. 4c-d shows a comparison of the $q$=0 EEL spectrum calculated with and without the long-range component of the Coulomb interaction for the 1L and bulk systems, respectively. Technically, the removal of the long-range screening is achieved by setting the divergent **G**=0 component of the Fourier transformed Coulomb potential, $V_G(\boldsymbol{q}) = 1/|\boldsymbol{G} + \boldsymbol{q}|^2$, to zero (here **G** denotes a reciprocal lattice vector). From Fig. 4c-d we conclude that in the optical limit, the monolayer response is not influenced by the long-range interactions while there is a very significant effect in the 3D bulk system. A direct consequence of this result is that for 2D materials the $q$=0 EEL spectrum coincides with the optical spectrum. It is well known that excitonic effects strongly influence the optical spectrum of a 2D semiconductor and thus the same must be expected for the $q$=0 EEL spectrum, and indeed this is the case (see discussion about the A peak below). However, since the RPA does not account for excitonic effects, the RPA spectrum of a 2D material in the $q$=0 limit simply reduces to the single-particle spectrum renormalized by the local field effects created by screening from the $\boldsymbol{G} \neq \boldsymbol{0}$ components of the Coulomb potential.

As previously noted the B and C plasmons are of interband nature. We now address the question of which bands/orbitals are involved in those plasmons. For simplicity we focus on the monolayer in the small $q$ limit where long range Coulomb interactions are negligible making the connection between the single-particle transitions and the spectral peaks straightforward. To make this identification, we note from Fig. 4c that the B and C plasmons originate from the two single-particle interband peaks (van Hove singularities) at about 4 eV and 8 eV, respectively. At finite $q$ the Coulomb interaction blue shifts these peaks creating the B and C plasmons. By considering the projected density of states (PDOS) we infer that the transitions involved in the B peak are from a lower valence band of mainly Pt-*d* character situated approximately 2 eV below the to the lowest conduction band with mainly Se-*p* orbitals. Similarly, the C peak originates from transitions from Pt-*p* to Pt-*s* orbitals, see Fig. S4.

**Excitonic nature of the A peak**

We now turn to the origin of the A peak that we have provisionally referred to as an excitonic peak. To support this interpretation, we perform a BSE calculation for the monolayer in the optical limit. Due to the well-known tendency of PBE to underestimate band gaps, we use the $G_0W_0$-approximation to correct the band gap of the PBE band structure[35, 36, 37, 38]. A comparison of the RPA and BSE results can be seen in Fig. 5a. For completeness we show the RPA result based on the PBE energies and the $G_0W_0$-band gap corrected PBE energies, respectively. The large difference between the BSE and RPA spectra highlights the importance of the direct e-h interaction, which drives the formation of excitons and leads to a dramatic rearrangement of spectral weights. We obtain a binding energy of the lowest A exciton of 0.51 eV, which compares well to previous calculations[22, 23, 24]. Because the $G_0W_0$-correction of the PBE band gap is comparable in size to the exciton binding energy, the peak in the RPA-PBE spectrum around 2 eV (which essentially is a peak in the joint density of states of the PBE single-particle spectrum) coincides almost with the lowest excitonic peak in the BSE-$G_0W_0$ spectrum. This explains the relatively good agreement between the RPA-PBE spectra and the experimental spectra in Figure 2a,d.

Next, we analyse the wave function of the A exciton as obtained from the BSE calculation of the PtSe$_2$ monolayer. Figure 5b shows the distribution of expansion coefficients for the e-h pairs forming the lowest exciton, on the band structure and the Brillouin zone (BZ) (inset), respectively. It appears that the exciton is made up of transitions from the highest valence band to the lowest conduction band located at the midpoint of Γ and M in the BZ.

Due to the large number of atoms in the 4L structure, it is difficult to perform accurate BSE calculations for this system. The fact that the A peak is only visible in the experimental EEL spectrum for small $q$ could indicate, according to the analysis presented earlier, that the A peak is also of excitonic origin in the 4L structure. On the other hand, the existence of excitonic states in the 4L structure seems unlikely due to its metallic nature, which intuitively should screen the direct e-h interaction and prevent exciton formation. However, recalling that screening is strongly suppressed in 2D materials in the small-$q$ limit, it cannot be ruled out that the e-h interaction is sufficiently attractive that excitons may form in the optical limit despite the metallic band structure, and that the A peak in the 4L structure could be of excitonic origin.

We performed a BSE calculation for bulk $PtSe_2$ to investigate how the inclusion of the direct e-h interaction affects the loss spectrum. As shown in Fig. S5, we find no significant excitonic effects in the bulk system that can account for the presence of the A-peak. This is as one would expect for a 3D metallic system where excitonic effects should indeed be screened out. We speculate that the A-peak observed in the experimental spectrum of a bulk system for small $q$ could represent an intraband plasmon. We obtain an in-plane (intraband) plasma frequency of 1.6 eV, which agrees well with the experimental A-peak position. However, in the full RPA spectrum the intraband plasmon peak is renormalized to 0.56 eV due to the screening by interband transitions. It should, however, be noted that the PBE functional is known to underestimate interband transition energies, which generally leads to an overestimation of interband screening and thus a red-shifting of plasmon energies in general[39]. However, we have not been able to confirm this effect from our calculations, and consequently the origin of the A-peak in the bulk system remains an open question.

## Discussion

In conclusion, we have unraveled the elementary electronic excitations in layered $PtSe_2$ using a combination of $q$-EELS measurements and theoretical calculations. Our analysis shows that the excitation spectrum, from monolayer to bulk, is governed by three distinct collective excitations, namely a low energy peak around 2 eV and two plasmons of 4-5 eV and 7-10 eV, respectively. The qualitatively different wave vector dependence of the dielectric function in 2D and 3D is manifested directly in the excitation spectra. In the monolayer and 4 layer samples, the low energy peak represents an exciton and completely dominates the excitation spectrum in the small $q$ limit, and only at larger $q$ do the plasmons develop and eventually dwarf the excitonic feature. In the bulk, the low energy peak and the plasmons exist side by side even in the small $q$ limit. For all material thicknesses, the EEL spectrum becomes dominated by the plasmons as $q$ is increased beyond ~ 0.05 Å$^{-1}$. Our work advances the understanding of the connection between dielectric screening and the formation of collective excitations in solids, and establishes the fundamental basis for photonic and optoelectronic applications of low-dimensional PtSe$_2$.

## METHODS

**STEM imaging and $q$-EELS.** Atomically resolved ADF-STEM image was obtained on an aberration corrected scanning TEM (Triple C1, JEOL JEM-2100CF) operated at 60 kV. The convergence angle of the STEM beam was set as 35 mrad and acceptance angle 70~200 mrad, and beam current was set to 15 pA. The $q$-EELS data was collected in diffraction mode using parallel beam illumination on a double Wiener Filtered monochromated TEM (Triple C2, JEOL ARM-60MΔ$^2$) operated at 60 kV, with an energy resolution of 40 ~ 50 meV and momentum resolution of 0.03 Å$^{-1}$. The illumination area of the samples for a selected area electron diffraction is approximately 300nm in diameter. Dual EELS mode of the Gatan Quantum 965 spectrometer was used to obtain the low loss spectra and zero loss peak (ZLP) simultaneously, to eliminate the effect of the ZLP drift during the spectrum acquisition. The dispersion of the spectrum was set to 0.005 eV/channel.

**DFT calculation.** The calculations were performed using the ASR framework[40] in combination with the GPAW electronic structure code[41]. To obtain the multilayer structures, we started from the relaxed T-phase PtSe2 monolayer (space group P-3m1(164)) structure obtained from C2DB[22, 42], obtaining the N-layer structures by stacking N monolayers in the most favorable AA' stacking and relaxed it using the PBE-D3 scheme[43] to account for the vdW-interaction between the layers. We performed the ground state calculations for the monolayer, multilayer and bulk systems using the asr.gs recipe. The calculations were performed in plane wave mode using a double zeta polarized basis set, an energy cut-off of 800 eV and a gamma-point centered k-point grid with a density of 12 k-points/Å$^2$ for the multilayer systems and a k-point density of 12 k-points/Å$^3$ for the bulk system. From the ground state calculations, we can then get the band structure using the ASR band structure recipe. The spin-orbit coupling is added non-self-consistently.

**RPA and BSE calculations.** To perform the EELS calculations within the RPA, we extended the ASR framework interface with the GPAW code linear response module to enable EELS calculations. Moving forward, the EELS module will be available in ASR. To ensure converged results, we interpolate the ground state calculation onto a k-point grid with an in-plane density of 50 k-points/Å$^2$ and include conduction bands equal to five times the number of valence bands. We further employ a broadening of 50 meV, and a local field cut-off of 50 eV. We employ a nonlinear frequency grid with domega_0 = 0.005 eV using the tetrahedron integration method for the mono- and multilayer systems, while employing the normal integration method for the bulk system.

The BSE calculations were performed on a k-point grid with an in-plane density of 6 k-points/Å$^2$. We included the four top valence bands and the lowest four conduction bands in the calculation of the BSE Hamiltonian. The calculation of the electronic screening was performed using 84 bands, all of the conduction bands equal to five times the number of valence bands, with a local field cut-off of 50 eV. The EELS spectrum was then calculated on a 10001-point frequency grid spanning 0-8 eV. To get the $G_0W_0$ correction to the quasiparticle energies we used the asr.gw recipe. To generate the excitonic weights plot for the lowest lying exciton, we extracted the excitonic weights from the BSE calculation and then matched the k-points of the band structure calculation to the k-point grid of the BSE calculation to get the projection.

## Data availability

All data that support the findings of this study have been included in the main text and Supplementary Information. Any additional materials and data are available from the corresponding author upon reasonable request.

## Acknowledgements

This work was supported by JST-CREST (JPMJCR20B1, JPMJCR20B5, JPMJCR1993) and JSPS-KAKENHI (JP16H06333, JP19K04434 and JP17H04797) and A3 Foresight program. H. Xu acknowledges the support from the National Natural Science Foundation of China (51972204). The Center for Nanostructured Graphene (CNG) is sponsored by the Danish National Research Foundation, Project DNRF103. This project has received funding from the European Research Council (ERC) under the European Union's Horizon 2020 research and innovation program grant agreements No 773122 (LIMA) and No 951215 (MORE-TEM). K. S. T. is a Villum Investigator supported by VILLUM FONDEN (grant no. 37789).

## Author Contributions

J.H. and K.S. conceived this project. J.H. performed the $q$-EELS measurement; M.S. and K.T. did the theoretical calculation; H. X. contributed the samples; J.H., M.S., M.K., K.T., K.S., T.P. commented and analyzed the data and theory; all authors co-wrote the paper with full discussion.

## Competing interests

The authors declare no competing interests.

## Additional information

**Supplementary information** The supplementary material is available online at

**Correspondence and requests for materials** should be addressed to M.S. or K.Z.

**Reprints and permissions information** is available at www.nature.com/reprints.

## References

1. Chhowalla, M., *et al.* The chemistry of two-dimensional layered transition metal dichalcogenide nanosheets. *Nat. Chem.* **5**, 263-275 (2013).

2. Wang, Q. H., Kalantar-Zadeh, K., Kis, A., Coleman, J. N., Strano, M. S. Electronics and optoelectronics of two-dimensional transition metal dichalcogenides. *Nat. Nanotechnol.* **7**, 699-712 (2012).

3. Wang, Y., *et al.* Monolayer PtSe$_2$, a new semiconducting transition-metal-dichalcogenide,

epitaxially grown by direct selenization of Pt. *Nano Lett.* **15**, 4013-4018 (2015).

4. Zhao, Y., *et al.* High-electron-mobility and air-stable 2D layered PtSe$_2$ FETs. *Adv. Mater.* **29**, 1604230 (2017).

5. Zhao, Y., *et al.* Extraordinarily strong interlayer interaction in 2D layered PtS$_2$. *Adv. Mater.* **28**, 2399-2407 (2016).

6. Ciarrocchi, A., Avsar, A., Ovchinnikov, D., Kis, A. Thickness-modulated metal-to-semiconductor transformation in a transition metal dichalcogenide. *Nat. Commun.* **9**, 919 (2018).

7. Chia, X., *et al.* Layered platinum dichalcogenides (PtS$_2$, PtSe$_2$, and PtTe$_2$) electrocatalysis: monotonic dependence on the chalcogen size. *Adv. Funct. Mater.* **26**, 4306-4318 (2016).

8. Lin, S., *et al.* Tunable active edge sites in PtSe$_2$ films towards hydrogen evolution reaction. *Nano Energy* **42**, 26-33 (2017).

9. Avsar, A., *et al.* Defect induced, layer-modulated magnetism in ultrathin metallic PtSe$_2$. *Nat. Nanotechnol.* **14**, 674-678 (2019).

10. Avsar, A., *et al.* Probing magnetism in atomically thin semiconducting PtSe$_2$. *Nat. Commun.* **11**, 4806 (2020).

11. Xu, H., *et al.* Controlled doping of wafer-scale PtSe$_2$ films for device application. *Adv. Funct. Mater.* **29**, 1805614 (2019).

12. Yu, X., *et al.* Atomically thin noble metal dichalcogenide: a broadband mid-infrared semiconductor. *Nat. Commun.* **9**, 1545 (2018).

13. García de Abajo, F. J. Optical excitations in electron microscopy. *Rev. Mod. Phys.* **82**, 209-275 (2010).

14. Koppens, F. H. L., Chang, D. E., García de Abajo, F. J. Graphene plasmonics: a platform for strong light–matter interactions. *Nano Lett.* **11**, 3370-3377 (2011).

15. Lourenço-Martins, H., Gérard, D., Kociak, M. Optical polarization analogue in free electron beams. *Nat. Phys.* **17**, 598-603 (2021).

16. Wang, G., Wang, Z., McEvoy, N., Fan, P., Blau, W. J. Layered PtSe$_2$ for sensing, photonic, and (opto-)electronic applications. *Adv. Mater.* **33**, 2004070 (2021).

17. Zeng, L., *et al.* Ultrafast and sensitive photodetector based on a PtSe$_2$/silicon nanowire array heterojunction with a multiband spectral response from 200 to 1550 nm. *NPG Asia Mater.* **10**, 352-362 (2018).

18. Wang, L., *et al.* Nonlinear optical signatures of the transition from semiconductor to semimetal in PtSe$_2$. *Laser & Photonics Reviews* **13**, 1900052 (2019).

19. Jakhar, A., Kumar, P., Moudgil, A., Dhyani, V., Das, S. Optically pumped broadband terahertz modulator based on nanostructured PtSe$_2$ thin films. *Adv. Opt. Mater.* **8**, 1901714 (2020).

20. Chernikov, A., *et al.* Exciton binding energy and nonhydrogenic Rydberg series in monolayer WS$_2$. *Phys. Rev. Lett.* **113**, 076802 (2014).

21. Ugeda, M. M., *et al.* Giant bandgap renormalization and excitonic effects in a monolayer transition metal dichalcogenide semiconductor. *Nat. Mater.* **13**, 1091-1095 (2014).

22. Haastrup, S., *et al.* The computational 2D materials database: high-throughput modeling and discovery of atomically thin crystals. *2D Mater.* **5**, 042002 (2018).

23. Thygesen, K. S. Calculating excitons, plasmons, and quasiparticles in 2D materials and van der Waals heterostructures. *2D Mater.* **4**, 022004 (2017).

24. Sajjad, M., Singh, N., Schwingenschlögl, U. Strongly bound excitons in monolayer PtS$_2$ and PtSe$_2$. *Appl. Phys. Lett.* **112**, 043101 (2018).

25. Bae, S., *et al.* Exciton-dominated ultrafast optical response in atomically thin PtSe$_2$. *Small* **17**, 2103400 (2021).

26. Ghosh, B., *et al.* Broadband excitation spectrum of bulk crystals and thin layers of PtTe$_2$. *Phys. Rev. B* **99**, 045414 (2019).

27. Habenicht, C., Knupfer, M., Büchner, B. Investigation of the dispersion and the effective masses of excitons in bulk 2H−MoS$_2$ using transition electron energy-loss spectroscopy. *Phys. Rev. B* **91**, 245203 (2015).

28. Kramberger, C., *et al.* Linear plasmon dispersion in single-wall carbon nanotubes and the collective excitation spectrum of graphene. *Phys. Rev. Lett.* **100**, 196803 (2008).

29. Wachsmuth, P., *et al.* High-energy collective electronic excitations in free-standing single-layer graphene. *Phys. Rev. B* **88**, 075433 (2013).

30. Roth, F., König, A., Fink, J., Büchner, B., Knupfer, M. Electron energy-loss spectroscopy: a versatile tool for the investigations of plasmonic excitations. *Journal of Electron Spectroscopy and Related Phenomena* **195**, 85-95 (2014).

31. Onida, G., Reining, L., Rubio, A. Electronic excitations: density-functional versus many-body Green's-function approaches. *Rev. Mod. Phys.* **74**, 601-659 (2002).

32. Onida, G., Reining, L., Godby, R. W., Del Sole, R., Andreoni, W. *Ab initio* calculations of the quasiparticle and absorption spectra of clusters: the sodium tetramer. *Phys. Rev. Lett.* **75**, 818-821 (1995).

33. Rohlfing, M., Louie, S. G. Electron-hole excitations in semiconductors and insulators. *Phys. Rev. Lett.* **81**, 2312-2315 (1998).

34. Andersen, K., Jacobsen, K. W., Thygesen, K. S. Spatially resolved quantum plasmon modes in metallic nano-films from first-principles. *Phys. Rev. B* **86**, 245129 (2012).

35. Hedin, L. New method for calculating the one-particle Green's function with application to the electron-gas problem. *Phys. Rev.* **139**, A796-A823 (1965).

36. Hybertsen, M. S., Louie, S. G. Electron correlation in semiconductors and insulators: band gaps and quasiparticle energies. *Phys. Rev. B* **34**, 5390-5413 (1986).

37. Aulbur, W. G., Jönsson, L., Wilkins, J. W. Quasiparticle calculations in solids. In: Ehrenreich H, Spaepen F (eds). *Solid State Physics*, vol. 54. Academic Press, 2000, pp 1-218.

38. Golze, D., Dvorak, M., Rinke, P. The GW compendium: a practical guide to theoretical photoemission spectroscopy. *Frontiers in Chemistry* **7**, 377 (2019).

39. Yan, J., Jacobsen, K. W., Thygesen, K. S. Conventional and acoustic surface plasmons on noble metal surfaces: A time-dependent density functional theory study. *Phys. Rev. B* **86**, 241404 (2012).

40. Gjerding, M.*, et al.* Atomic simulation recipes: A Python framework and library for automated workflows. *Comput. Mater. Sci.* **199**, 110731 (2021).

41. Enkovaara, J.*, et al.* Electronic structure calculations with GPAW: a real-space implementation of the projector augmented-wave method. *J. Phys. Condens. Mat.* **22**, 253202 (2010).

42. Gjerding, M. N., et al. Recent progress of the computational 2D materials database (C2DB). *2D Mater.* **8**, 044002 (2021).

43. Grimme, S., Antony, J., Ehrlich, S., Krieg, H. A consistent and accurate ab initio parametrization of density functional dispersion correction (DFT-D) for the 94 elements H-Pu. *J. Chem. Phys.* **132**, 154104 (2010).

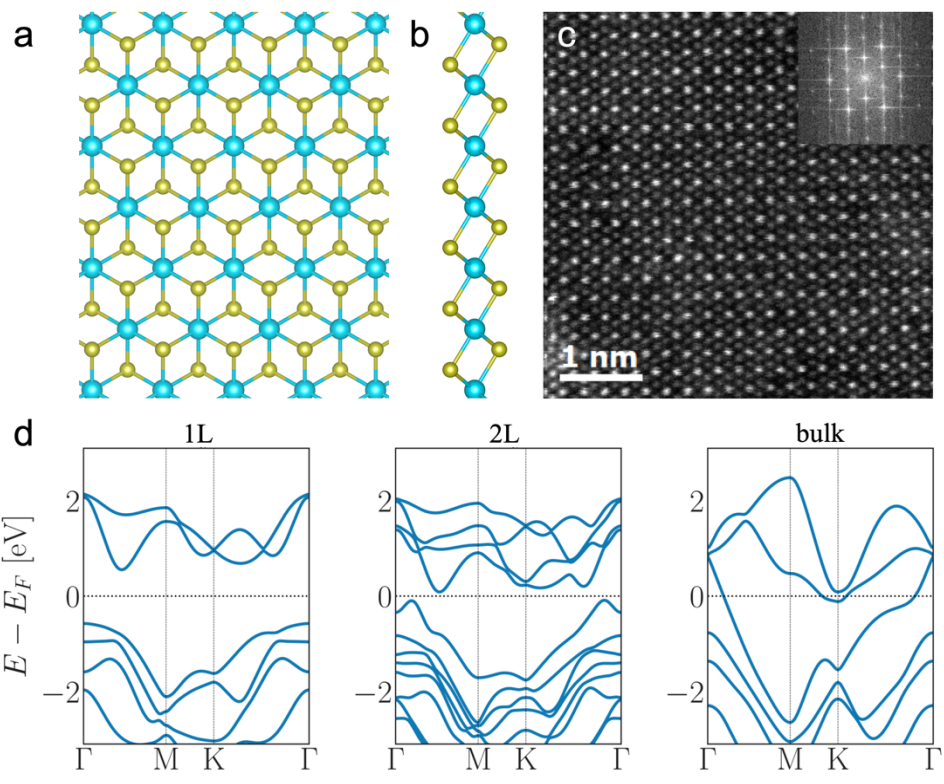

**Fig. 1 | Atomic and electronic structures of PtSe$_2$. a-b,** Top- and side-view structure models of monolayer 1TPtSe$_2$. Cyan balls are Pt and yellow are Se atoms. **c,** ADF-STEM image of monolayer PtSe$_2$. **d,** Thickness dependent electronic band structure of PtSe$_2$ in the 2D Brillouin zone.

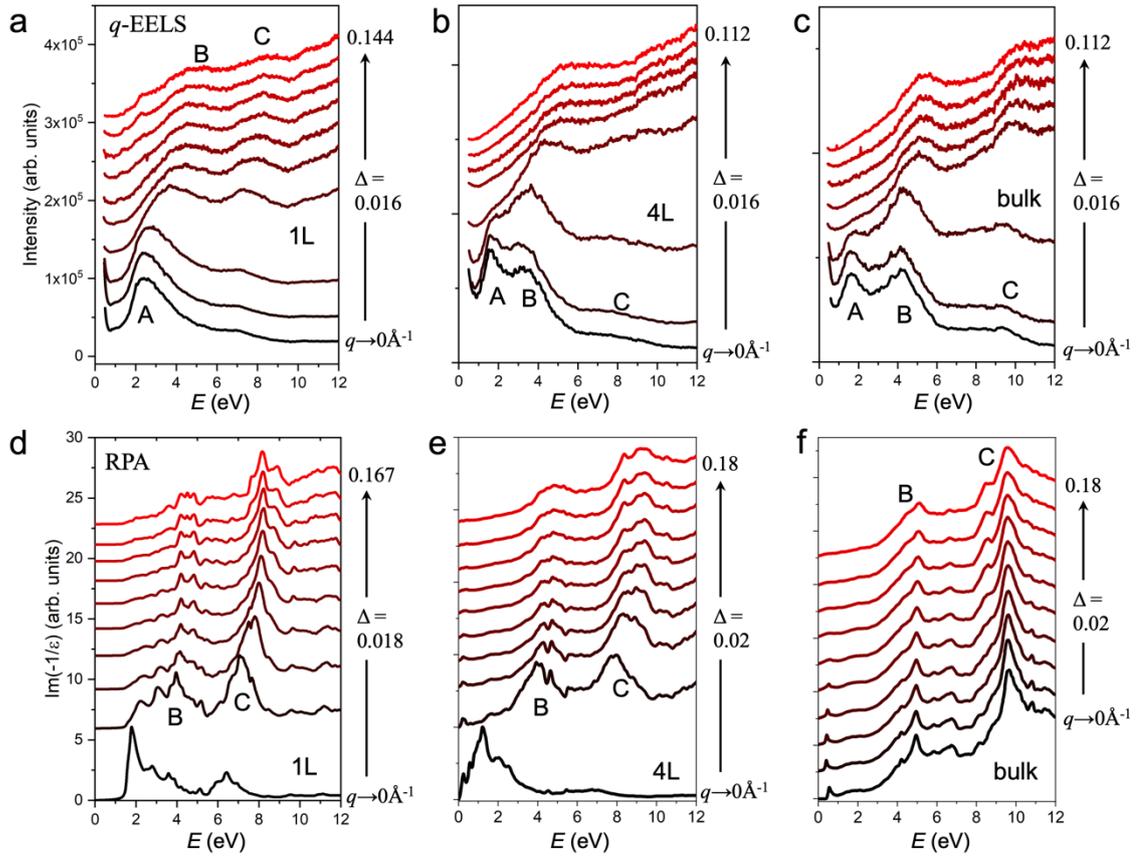

**Fig. 2 | Momentum-dependent loss function of PtSe$_2$. a-c,** Experimental $q$-EELS of 1L, 4L and bulk PtSe$_2$, respectively. **d-f,** Calculated $q$-dependent loss function in the random phase approximation (RPA) of 1L, 4L and bulk PtSe$_2$, respectively. The low energy excitonic peak A, and the intermediate and high energy plasmonic peaks B and C are indicated.

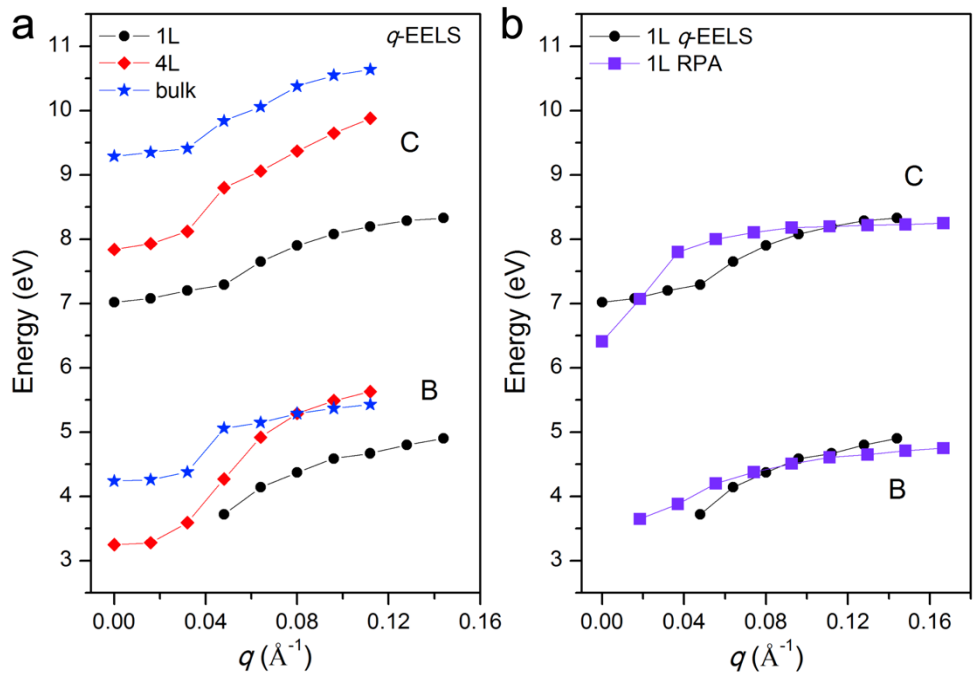

**Fig. 3 | Dispersion of the loss peaks of PtSe$_2$. a,** Experimental dispersion of the B and C peaks of PtSe$_2$ for different thicknesses. **b,** Comparison of the experimental and RPA calculated dispersion of the B and C peaks for monolayer PtSe$_2$.

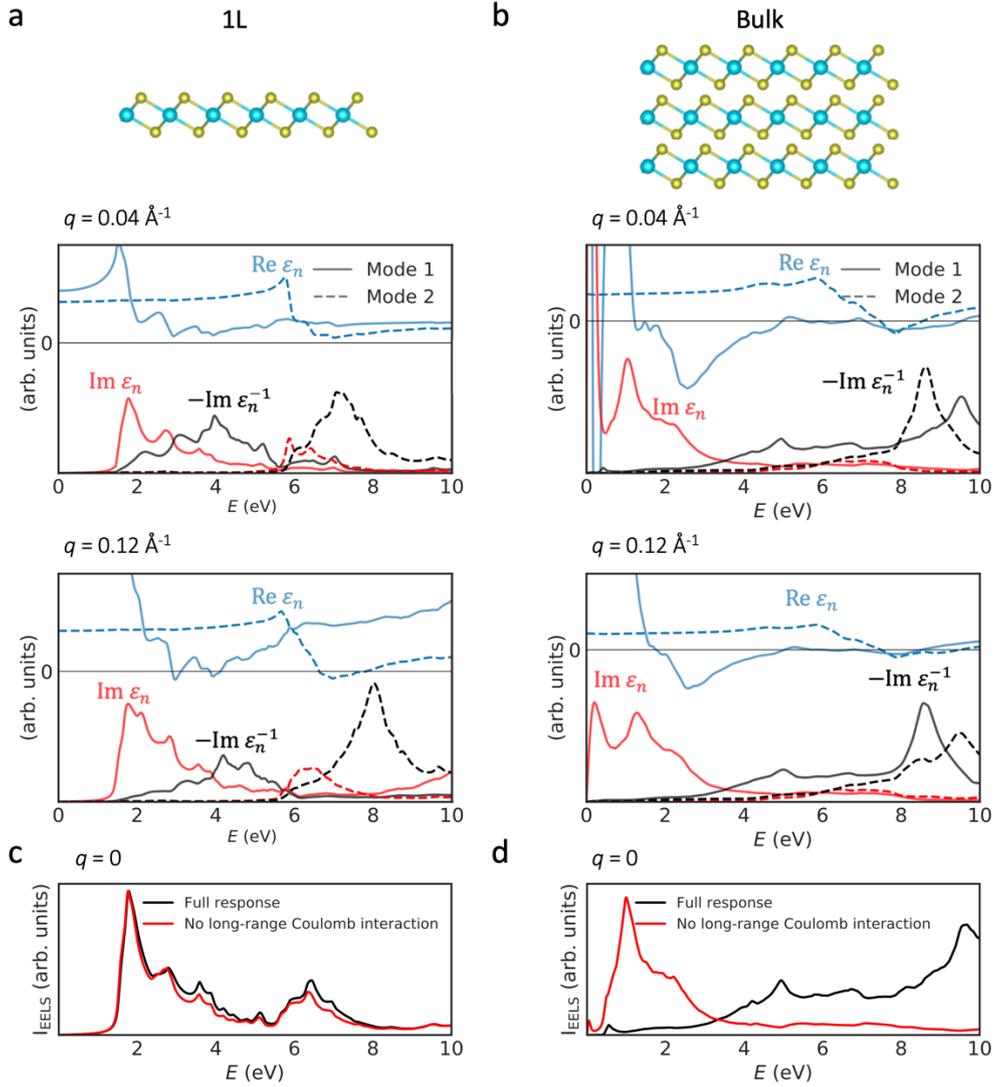

**Fig. 4 | Plasmon formation in 2D vs 3D. a-b,** Evolution of the eigenvalues obtained from the spectral decomposition of the RPA dielectric function for 1L and bulk PtSe$_2$ as a function of $q$. The loss spectra of the 1L system are more sensitive to $q$ than those of the bulk because of the gradual turning on of the long-range screening in 2D with increasing $q$. **c-d,** The macroscopic RPA loss function in the $q$=0 limit for 1L and bulk. Inclusion of the long-range component of the Coulomb interaction (the Fourier component corresponding reciprocal lattice vector $G$=0) has essentially no effect on the 1L spectrum while the bulk spectrum is affected dramatically due to the formation of plasmons.

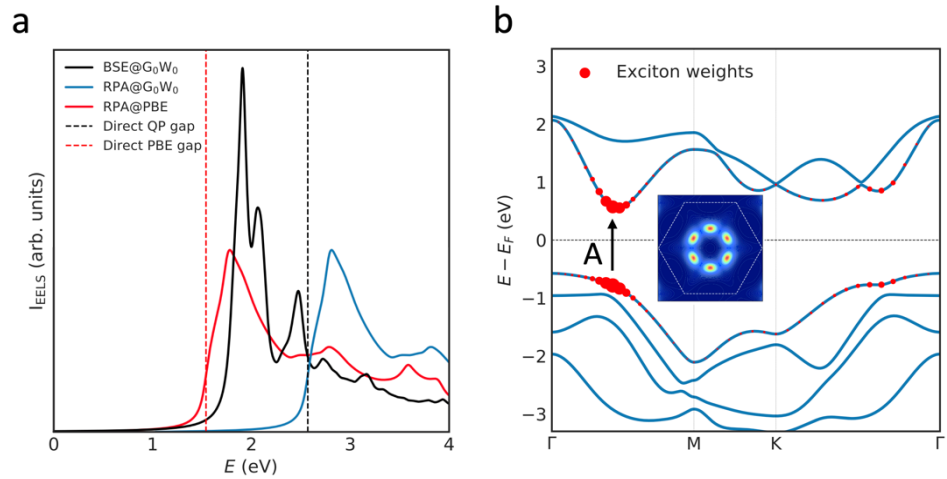

**Fig. 5 | Excitonic origin of the A-peak. a,** Comparison of BSE and RPA calculated loss functions of monolayer PtSe$_2$ in the $q \rightarrow 0$ limit. The BSE spectrum is calculated on top of the G$_0$W$_0$ band structure. For comparison, the RPA spectrum is calculated on top of the PBE band structure and the PBE band structure corrected to match the G$_0$W$_0$ band gap, respectively. **b,** The BSE calculated excitonic weights of the A peak. The colored inset shows the 2D projected exciton wavefunction distribution in $k$ space. Both the excitonic weights and $k$-space wavefunction show that the A exciton originates from direct transitions at the midpoint of ΓM (highlighted by the black arrow).